# Selective Fermi-Level Pinning: A Design Strategy for Giant Rectification in Molecular Junctions

*Junnan Guo[1], Wenhui Fang[1], Jian Huang[1], Weikang Wu[1*], Hui Li[1*], and Lishu Zhang[1*],*

[1]Key Laboratory for Liquid-Solid Structural Evolution and Processing of Materials, Ministry of Education, Shandong University, Jinan 250061, China

E-mail: weikang_wu@sdu.edu.cn

E-mail: lihuilmy@hotmail.com

E-mail: lishu.zhang@sdu.edu.cn

## Abstract

Molecular rectifiers are key functional components of molecular-scale integrated circuits, yet achieving high rectification ratios remains a longstanding challenge due to the intrinsic symmetry of resonant tunneling and the complexity of interfacial energy-level alignment. Here, we propose a rectifier design strategy based on selective Fermi-level pinning that breaks transport symmetry via pinning interactions between molecular frontier orbitals and electrodes. This framework enforces tunneling transport to be predominantly governed by unoccupied molecular orbitals, while substantially suppressing contributions from occupied states, thereby establishing a simplified and highly controllable rectification mechanism. The resulting cyclo[*n*]carbon-based molecular junctions exhibit giant rectification ratios exceeding $10^3$, while retaining exceptional structural robustness against variations in both donor chain length and carbon ring size. This work reveals the critical role of selective Fermi-level pinning in molecular junctions and provides a general design principle for engineering functional single-molecule electronic devices.

# 1. Introduction

As silicon-based integrated circuits approach their physical scaling limits, single-molecule devices have emerged as a promising route toward ultimate miniaturization.[1-3] Molecular rectifiers, as core functional units in molecular electronics, play indispensable roles in molecular-scale integrated circuits, where their performance directly determines the response speed and operational stability of the system.[4, 5]

The performance of molecular rectifiers is usually evaluated by the rectification ratio ($RR = \left|\frac{I^+}{I^-}\right|$).[6] Since the Aviram-Ratner model was first proposed, great deal of efforts have been devoted to molecular engineering, interfacial coupling optimization, quantum interference (QI), and external field modulation.[7-12] Despite these advances, high-performance rectification at the single-molecule level remains fundamentally constrained.[13-16] The primary limitation arises from the intrinsic symmetry of resonant tunneling and the complex evolution of energy level alignment under applied bias at molecule electrode interfaces.[17, 18] This evolution results from a strongly coupled interplay of interface push-back effects, Fermi-level pinning, charge transfer, and interfacial dipole reconstruction, which collectively hinder precise control of molecular orbital alignment relative to the electrode Fermi level and thus restrict device performance.[19-22]

Among these interfacial effects, Fermi-level pinning emerges as pivotal governing factor.[19, 23] It originates from strong interactions between interface states and electrode electronic states, which partially lock molecular energy levels near the electrode Fermi level and significantly diminish the efficacy of other tuning parameters. In traditional semiconductor contact systems, this effect is generally regarded as detrimental, as it limits Schottky barrier tunability.[24, 25] In molecular-scale systems, however, fermi-level pinning provides a novel physical dimension for interfacial energy-level manipulation.[26, 27] Previous studies have demonstrated that selective pinning of molecular orbitals can be achieved by engineering the chemical bonding configuration and local electronic structure at the molecular-electrode interface.[28, 29] When distinct molecular orbitals exhibit divergent pinning behaviors at the interface, orbital-dependent asymmetric electronic coupling emerges, which breaks transport symmetry in resonant tunneling processes and enhances rectification performance.

Here, we propose a universal design strategy for single-molecule rectifiers based on selective Fermi-level pinning, which is verified using cyclo[*n*]carbon-based molecular junctions as the model system. This strategy enables spatial separation of frontier molecular orbitals (FMOs) and strictly confines the resonant tunneling process to a single bias polarity through the synergistic integration of non-coplanar donor units, wide-bandgap cyclo[*n*]carbon acceptors and asymmetric anchoring groups. The resulting cyclo[*n*]carbon-based molecular junctions exhibit giant RRs exceeding $10^3$ and maintain excellent structural robustness against variations in donor chain length and carbon ring size. Further universality analysis reveals that the rectification behavior based on this strategy is jointly controlled by the electrode density of states (DOS) and the acceptor bandgap. This work reveals the critical role of selective Fermi-level pinning in regulating molecular rectification and provides a generalizable design principle for the rational engineering of sub-nanometer-scale functional molecular devices.

## 2. Computational methods

Device optimization and transport property calculations are performed using Quantum ATK based on the non-equilibrium Green's function combined with density functional theory (NEGF-DFT) method.[30] The initial geometries of the acceptor cyclo[*n*]carbon are optimized at the ωB97XD/def2-TZVP level, which provides reliable accuracy for π-conjugated systems and long-range dispersion interactions, showing good agreement with available experimental data.[31-34] For device-level structural optimization and transport calculations, the exchange-correlation interaction is treated within the generalized gradient approximation (GGA) using the Perdew-Burke-Ernzerhof (PBE) functional.[35] A double-ζ polarized (DZP) basis set is employed for all atoms. During structural relaxation, the force convergence criterion is set to 0.01 eV/Å. A real-space grid cutoff energy of 75 Hartree and an electronic temperature of 300 K are used to ensure numerical stability and self-consistency convergence.[36]

The current-voltage (I-V) characteristics are calculated using the Landauer-Büttiker formalism:[37]

$$I = \frac{2e}{h}\int_{-\infty}^{\infty} dE(T(E,V)(f_1(E) - f_2(E)))$$

where $E$ is the energy of incident electrons, $f_1\,(E)$ and $f_2\,(E)$ are the Fermi-Dirac distribution functions of the source and drain electrodes, respectively, $e$ is the electron charge, and $h$ is Planck's constant. $T(E,V)$ denotes the transmission function, given by:

$$T(E,V) = tr(\Gamma_1 G^R \Gamma_2 G^A)$$

where $G^R$ and $G^A$ are the retarded and advanced Green's functions of the scattering region, respectively, and $\Gamma_1$ and $\Gamma_2$ represent the coupling matrices between the scattering region and the left/right electrodes.

## 3. Results and discussion

### 3.1 Selective Fermi-Level Pinning Transport Principle

Molecular rectifiers typically consist of two electrodes (left electrode L and right electrode R), with one or more functional molecules sandwiched between them, as illustrated in **Fig. 1**a. By converting alternating input current into unidirectional direct current output, these single-molecule rectifiers can exhibit rectification behavior analogous to that of semiconductor diodes, characterized by a large current under one bias polarity and suppressed current under the opposite polarity (**Fig. 1**b). This asymmetric transport originates from tunneling transport processes at the molecule-electrode interface, where energy-level alignment plays a decisive role.

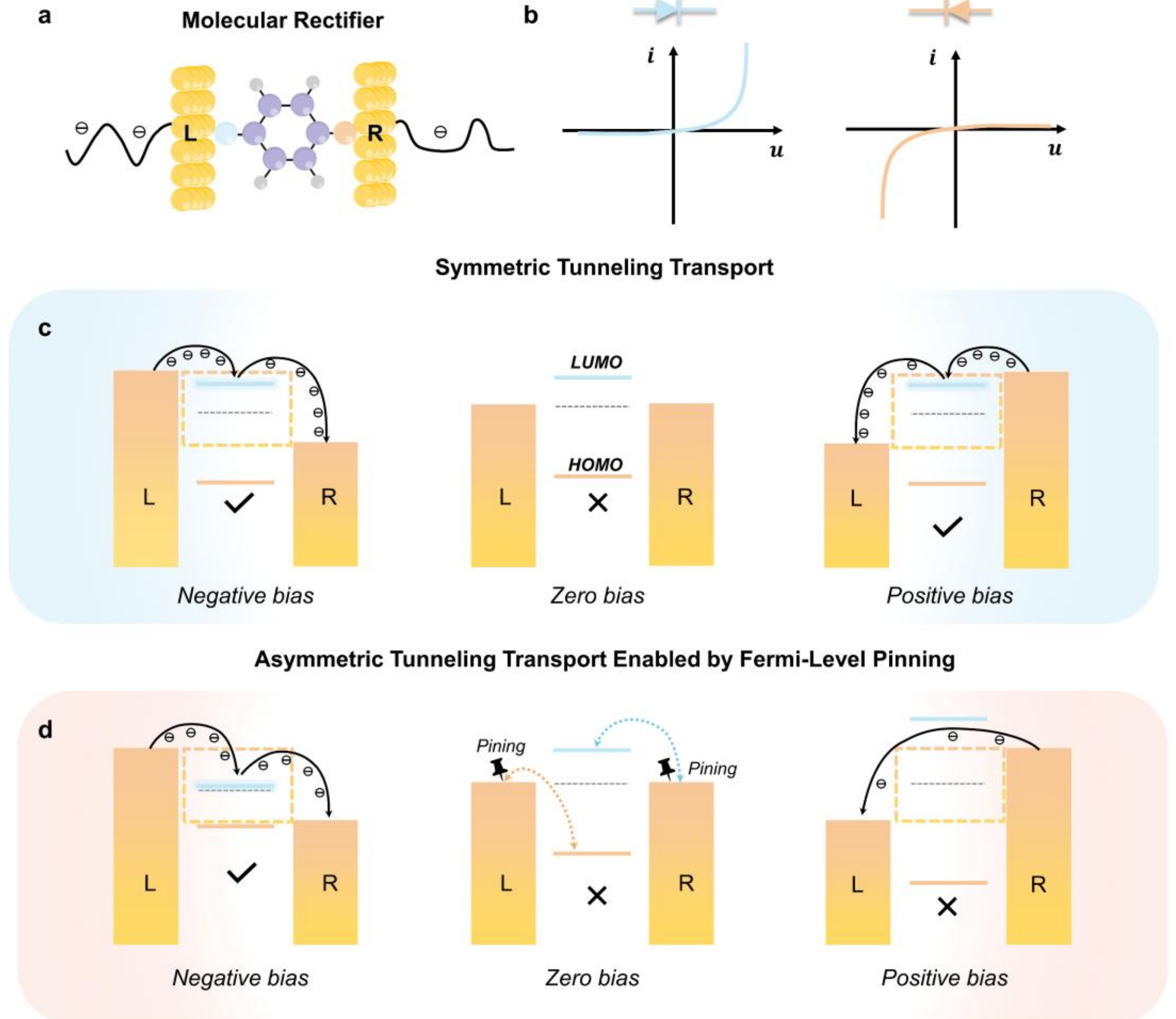

**Figure 1. Symmetric and asymmetric tunneling transport mechanisms. (a)** Schematic illustration of a molecular rectifier with a functional molecule sandwiched between the left (L) and right (R) electrodes. **(b)** Idealized current-voltage (I-V) characteristics demonstrating the rectification behavior. **(c)** Energy-level alignment of a typical molecular junction exhibiting symmetric tunneling transport. **(d)** Schematic of the asymmetric tunneling transport enabled by selective Fermi-level pinning.

In a typical molecular junction, interfacial charge redistribution establishes a unified Fermi level across the system (**Fig. 1**c). Under an applied bias, the electrochemical potentials of the two electrodes split, while electron transport proceeds through coherent tunneling, governed by the alignment between FMOs and the electrode Fermi level. When molecular orbitals enter the bias window, resonant tunneling occurs, leading to a high current; otherwise, transport is dominated by the off-resonant tunneling with significantly reduced current. In this framework, the electronic transport is intrinsically symmetric with respect to bias polarity, as molecular orbitals enter the bias window in a symmetric manner under opposite biases, making rectification challenging to achieve.

However, this symmetry can be broken through selective Fermi-level pinning, thereby enabling its rectification. As shown in **Fig. 1**d, when the lowest unoccupied molecular orbital (LUMO) is strongly coupled and pinned to the right electrode while the highest occupied molecular orbital (HOMO) is anchored to the left electrode, an asymmetric energy-level coupling configuration is constructed. Although the overall Fermi level remains between the two electrode chemical potentials, the frontier orbitals exhibit distinct responses under bias. Under one polarity, the pinned orbitals follow their respective electrode potentials but remain outside the bias window, and transport is dominated by off-resonant tunneling, resulting in off-resonant transport and low current. Under the opposite polarity, the LUMO is driven into the bias window, activating resonant tunneling and yielding a pronounced current increase. Consequently, tunneling transport becomes strongly bias-dependent, enabling rectification behavior.

### 3.2 Design Strategy and Transport Mechanism

To better achieve selective Fermi-level pinning and construct molecular rectifiers, we integrate the Van Dyck-Ratner framework with the donor-acceptor (D-A) paradigm and propose three synergistic design principles:[28, 38, 39]

(i) A non-coplanar long-chain donor is introduced to suppress intramolecular $\pi$-conjugation via structural torsion, thereby spatially separating the HOMO and LUMO.

(ii) A wide-bandgap acceptor is employed as the transport center is employed as the transport center to preserve a large energy gap after functionalization and D-A coupling, thereby suppressing the contribution from occupied frontier orbitals and restricting the tunneling transport to the unoccupied orbitals.

(iii) Asymmetric anchoring is implemented by functionalizing opposite molecular termini with electron-donating and electron-withdrawing groups, thereby enabling directional Fermi-level pinning.

Following these principles, 1,2-diphenylethane is selected as the non-coplanar donor, while cyclo[10]carbon serves as the wide-bandgap acceptor (**Fig. 2**a).[34, 40] Asymmetric terminal groups (–SH and –CN) are introduced to functionalize the two moieties and establish directional orbital-electrode coupling, forming a molecular junction with selective Fermi-level pinning characteristics, denoted as 2D-cyclo[10]carbon (**Fig. 2**b). Although the individual fragments (1,2-diphenylethane-SH and cyclo[10]carbon-CN) exhibit large bandgaps of 3.46 and 3.50 eV, respectively, the coupled D-A system shows a substantially reduced gap of 1.443 eV (**Fig. S1**a-c), indicating pronounced energy-level reconstruction and confirming the critical role of wide-bandgap building blocks. Au electrodes are selected due to their high DOS, which enhances the Fermi-level pinning effect. Upon 2D-cyclo[10]carbon junction formation, the frontier orbital alignment reveals that the LUMO lies close to the electrode Fermi level, while the HOMO is pushed far below it (**Fig. S1**c), establishing a LUMO-dominated transport configuration.

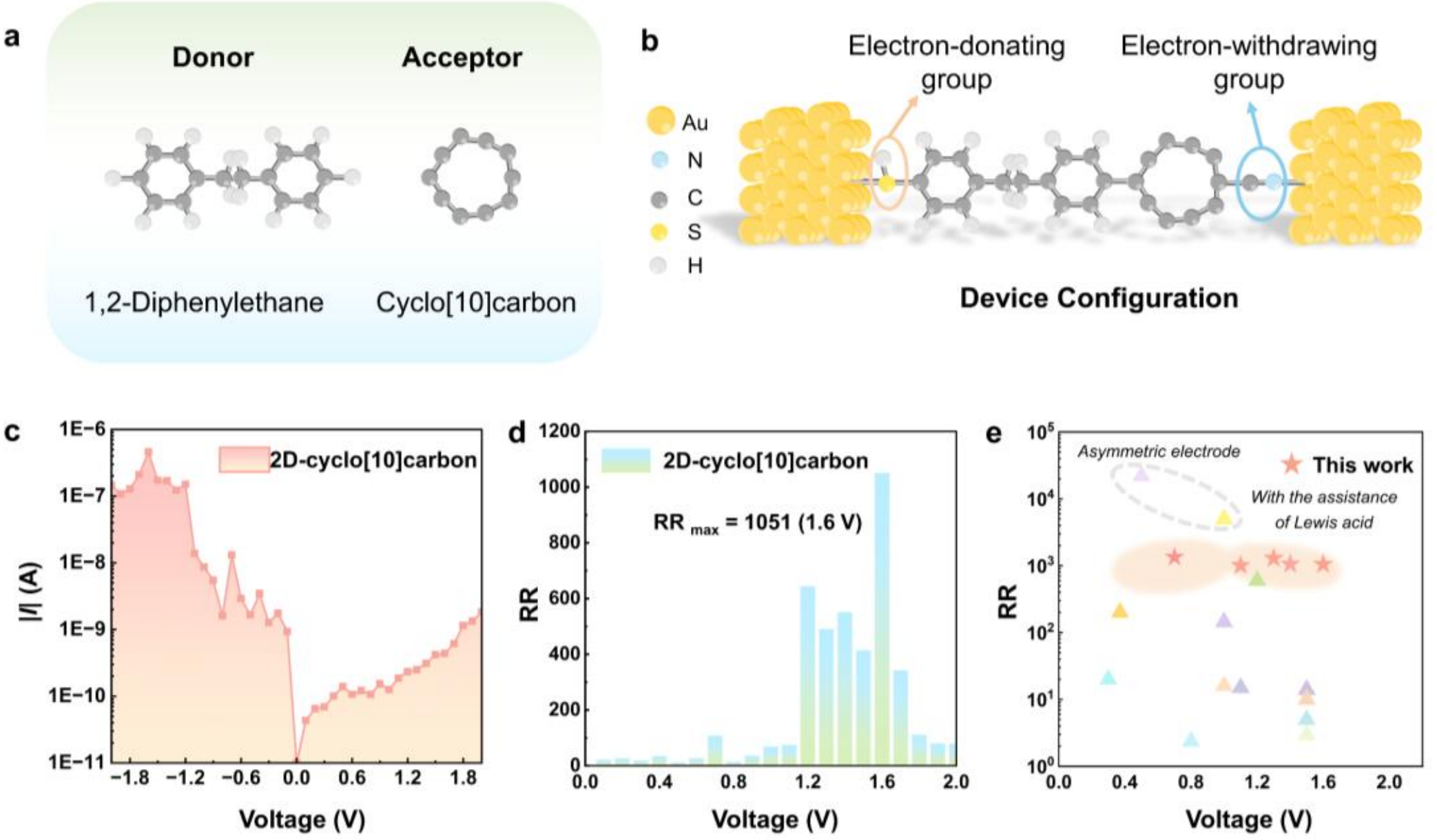

**Figure 2**. **Design strategy and rectification performance of cyclo[10]carbon-based junction.** **(a)** Optimized geometries of 1,2-diphenylethane and cyclo[10]carbon. **(b)** Device configuration of 2D-cyclo[10]carbon junction. **(c)** Corresponding I-V characteristics. **(d)** Bias-dependent rectification ratios (RRs) extracted from the I-V curves. **(e)** Summary of the RRs in this work compared with reported single-molecule rectifiers in recent studies. Corresponding references and values are summarized in **Table S1** of the Supplementary Information.

As shown in **Fig. 2**c, I-V characteristics of the molecular junction exhibits strongly asymmetric transport behavior, with a rapidly increasing current under negative bias and pronounced suppression under positive bias. The RR increases monotonically with bias and reaches a maximum value of 1051 at 1.6 V (**Fig. 2**d), achieving a high rectification performance exceeding $10^3$. Compared with previously reported single-molecule rectifiers (**Fig. 2**e), the present system shows clear advantages. Notably, while higher RRs in prior studies rely on asymmetric electrodes or Lewis acid-assisted modulation, the present device achieves superior performance solely through molecular-level design without external control, demonstrating that this strategy enables intrinsic regulation of charge transport at the molecular scale.

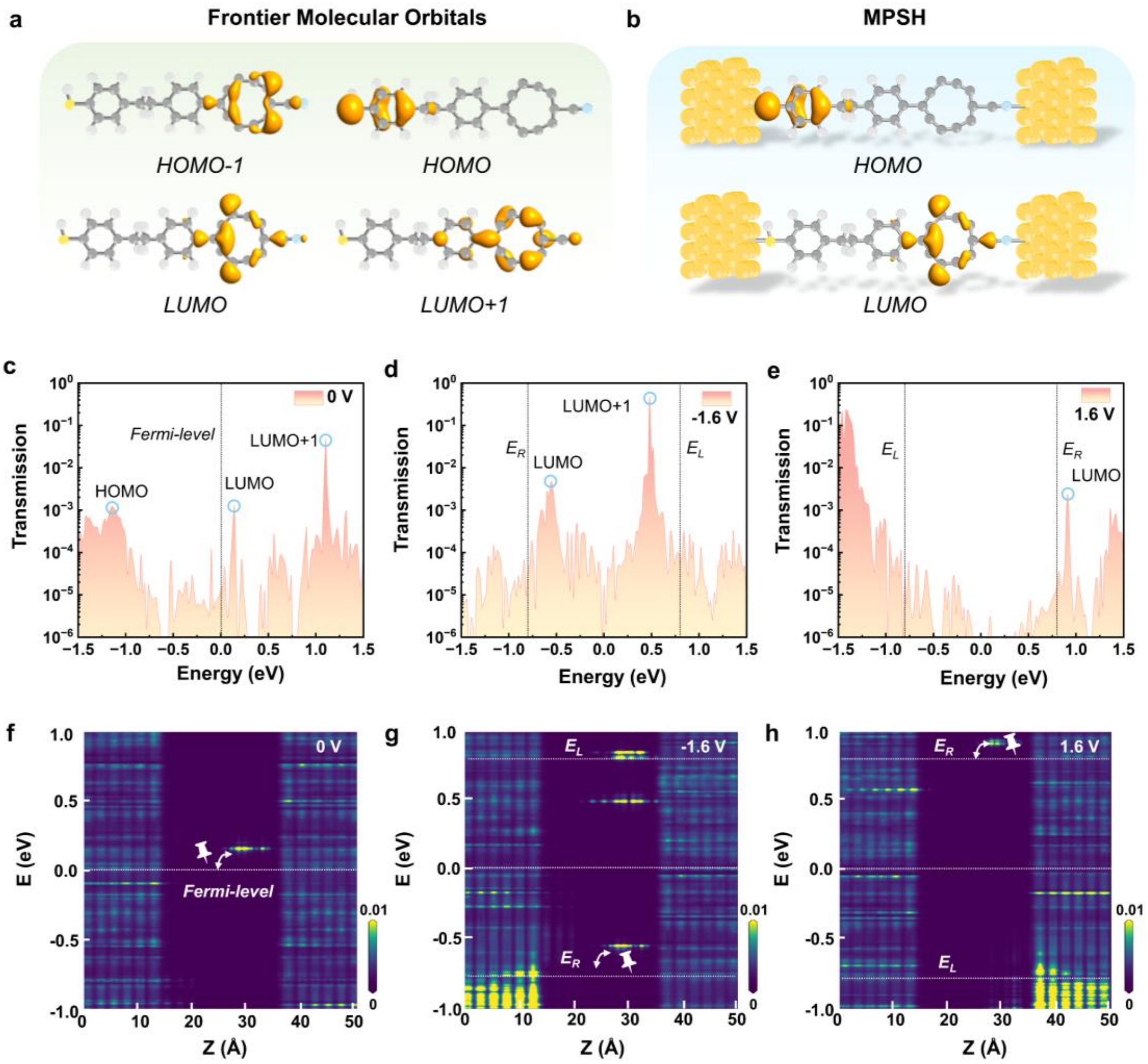

**Figure 3. Electronic structure and transport mechanism of 2D-cyclo[10]carbon junction. (a)** Frontier molecular orbitals (FMOs) of the 2D-cyclo[10]carbon molecule. **(b)** Molecule-projected self-consistent Hamiltonian (MPSH) of the highest occupied molecular orbital HOMO and the lowest unoccupied molecular orbital (LUMO) within the device. **(c-e)** Transmission spectra at (c) 0 V, (d) -1.6 V, and (e) 1.6 V. **(f-h)** Projected local density of states (PLDOS) at (f) 0 V, (g) -1.6 V, and (h) 1.6 V. The white solid lines indicate the bias window defined by the electrochemical potentials of the left ($E_L$) and right ($E_R$) electrodes.

To elucidate the microscopic origin of selective Fermi-level pinning and the resulting rectification behavior, we analyze the FMOs, molecular projected self-consistent Hamiltonian (MPSH), transmission spectra, and projected local density of states (PLDOS) (**Fig. 3**). As shown in **Fig. 3**a, the FMOs exhibit pronounced spatial separation. The HOMO is predominantly localized on the left donor unit, whereas the LUMO and LUMO+1 are mainly distributed over the right acceptor region. This spatial decoupling effectively suppresses inter-orbital hybridization and establishes the essential prerequisite for selective Fermi-level pinning. Upon formation of the molecular junction, MPSH analysis further confirms that the spatial distribution of the molecular orbitals remains largely preserved (**Fig. 3**b), providing a fundamental electronic basis for giant rectification performance. The bias-dependent evolution of the transmission spectra reveals the physical origin of the rectification behavior (**Fig. 3**c-e). The transmission peak at approximately 0.14 eV to the right of the Fermi level corresponds to the LUMO, while the peak at approximately 1.14 eV on the left side originates from the HOMO (**Fig. 3**c). Therefore, charge transport is primarily dominated by the LUMO channel. Meanwhile, the large HOMO-LUMO gap suppresses the contribution of occupied states near the transport window, leading to a simple transport mechanism. Under a negative bias of -1.6 V, the LUMO is driven into the bias window, giving rise to strong resonant tunneling and significantly enhanced electron transport (**Fig. 3**d). In contrast, under a positive bias of +1.6 V, all frontier orbitals remain outside the bias window, and transport is governed by off-resonant tunneling, resulting in strongly suppressed current (**Fig. 3**e). Moreover, the participation of LUMO+1 under negative bias further enhances the transmission intensity, thereby amplifying the reverse current and contributing to the ultrahigh RR. PLDOS analysis provides a spatial picture consistent with the transport results (**Fig. 3**f-h). At zero bias, the HOMO and LUMO are clearly localized at opposite ends of the molecule, maintaining a well-defined spatial separation that persists under external bias (**Fig. 3**f). Under negative bias, a continuous high intensity conduction pathway emerges along the right-side cyclo[10]carbon backbone, bridging the two electrodes and indicating the efficient resonant transport across the junction (**Fig. 3**g). In contrast, under positive bias, electronic states are largely confined near the electrodes, while the molecular region lacks an effective transport channel, leading to strongly hindered conduction (**Fig. 3**h). These findings collectively confirm that the giant rectification behavior originates from the asymmetric orbital response induced by selective Fermi-level pinning.

### 3.3 Geometric Effects and Mechanistic Origin

It is well known that variations in molecular structure can modulate intrinsic dipole moments and orbital distributions, thereby influencing rectification behavior. To evaluate the structural robustness of the selective Fermi-level pinning mechanism, we systematically investigate the effects of donor chain length and cyclo[n]carbon acceptor size on the transport characteristics (**Fig. 4**a).

The I-V characteristics and bias-dependent RRs (**Fig. 4**b-c) show that all devices exhibit pronounced current asymmetry and maintain high RRs, indicating strong tolerance against structural perturbations. When the cyclo[10]carbon acceptor is fixed, increasing the number of phenyl units from dimer (2D) to tetramer (4D) preserves high rectification performance, demonstrating that the selective pinning configuration remains robust. The maximum RR ($RR_{max}$) increases with molecular length, while the turn-on voltage decreases. In contrast, the absolute current decreases due to the exponential attenuation of coherent tunneling with increasing molecular length.[41] When the donor is fixed and the cyclo[n]carbon size is increased from $C_{10}$ to $C_{18}$, all systems retain RRs on the order of $10^3$, further demonstrating the generality of the design strategy. The bias corresponding to $RR_{max}$ shifts toward lower voltages with increasing ring size, while $RR_{max}$ exhibits a non-monotonic dependence, indicating a cooperative modulation of transport by the molecular structure and energy-level alignment.

Furthermore, transmission contour plots (**Fig. 4**d-f) reveal that variations in donor length do not alter the fundamental energy alignment. The electrode chemical potentials $\mu_L$ and $\mu_R$ are selectively pinned to donor- and acceptor-related states, respectively. Although a slight shift in the pinning distance between the LUMO and electrode Fermi level is observed with increasing cyclo[n]carbon size, the overall alignment remains essentially unchanged (**Fig. S**3). Meanwhile, the PLDOS results (**Fig. S**4 and **S**5) are consistent with the energy-level characteristics observed in transmission spectra, jointly verifying the stability of such transport features from both energy and spatial dimensions.

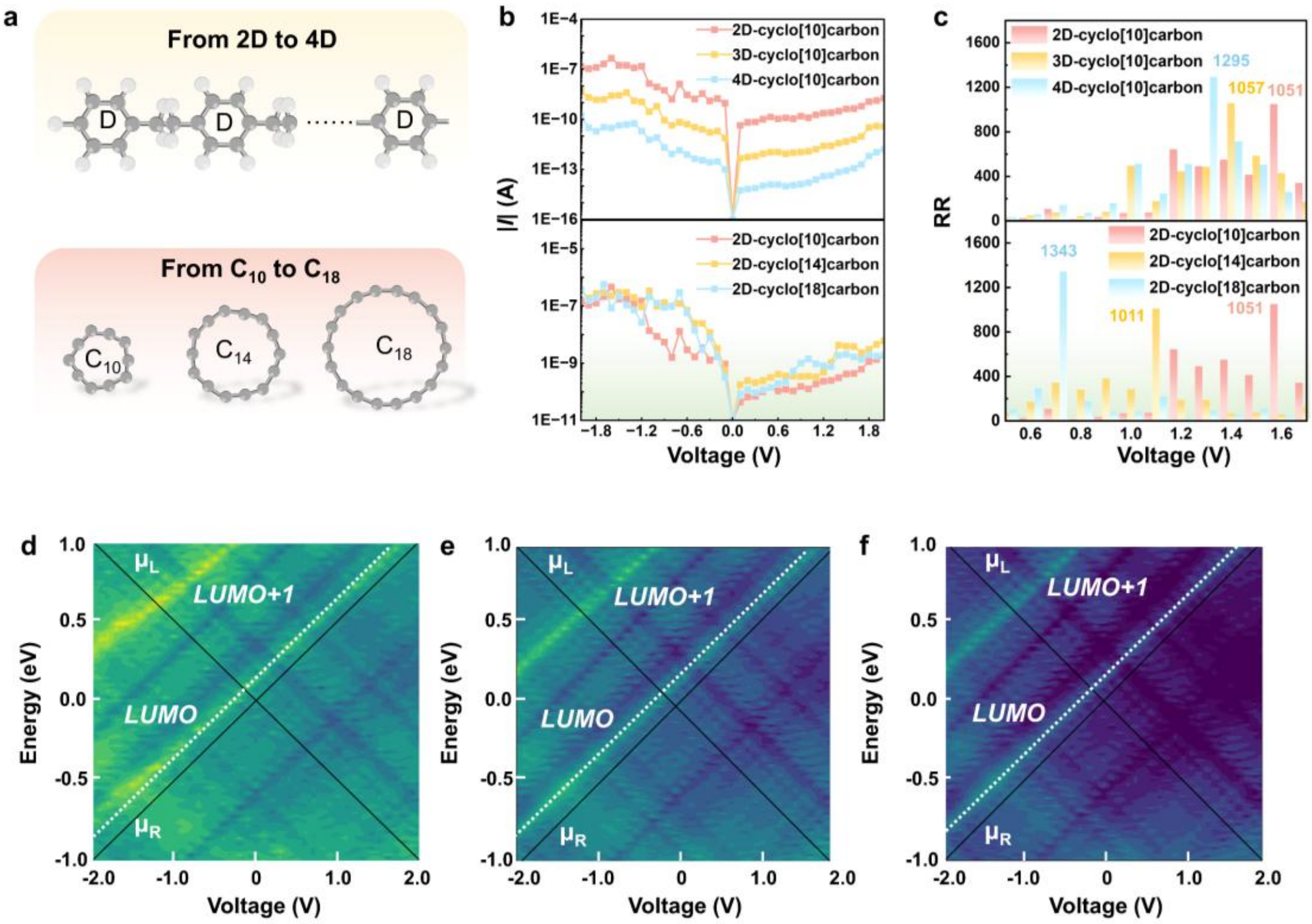

**Figure 4. Geometric Effects on rectification performance. (a)** Schematic of structural evolution from 2D (two phenyl rings) to 4D (four phenyl rings) donors and from cyclo[10]carbon to cyclo[18]carbon acceptors. **(b)** I-V characteristics of molecular junctions with varying donor lengths and acceptor ring sizes. **(c)** Bias-dependent RRs for the corresponding molecular junctions. **(d-f)** Contour plots of the transmission spectra for (d) 2D-cyclo[10]carbon, (e) 3D-cyclo[10]carbon, and (f) 4D-cyclo[10]carbon junctions. The white dashed lines track the LUMO energy shift, and black solid lines indicate the bias window.

To further elucidate the relationship between the molecular structure and rectification behavior, we analyze the local density of states (LDOS), electrostatic potential difference (EDP), and the device MPSH states of these junctions (**Fig. 5**). The LDOS results reveal that, for all systems, the electronic states are predominantly localized on the cyclo[*n*]carbon acceptor, while the donor units mainly act as structural linkers and contribute weakly to the electronic transport (**Fig. 5**a-e). In contrast, the EDP exhibits strong spatial asymmetry, with the potential drop concentrated on the donor side, generating a localized internal electric field. These observations indicate that the donor side provides an asymmetric electrostatic driving field, whereas the cyclo[n]carbon acceptor hosts the primary transport states. The rectification behavior arises from their cooperative effect. With increasing donor chain length, the LDOS distribution on the cyclo[n]carbon remains nearly unchanged, while the asymmetry of the EDP is progressively enhanced, accompanied by an expansion of the equipotential contour distribution (**Fig. 5**a-c). This facilitates the entry of the LUMO into the resonant transport regime at lower bias, leading to reduced turn-on voltage and enhanced RR. In contrast, enlarging the cyclo[n]carbon size leads to stable EDP but distinct variations in LDOS (**Fig. 5**a, d and e), indicating that transport gradually shifts from electrostatic-field-dominated to electronic-structure-dominated behavior.

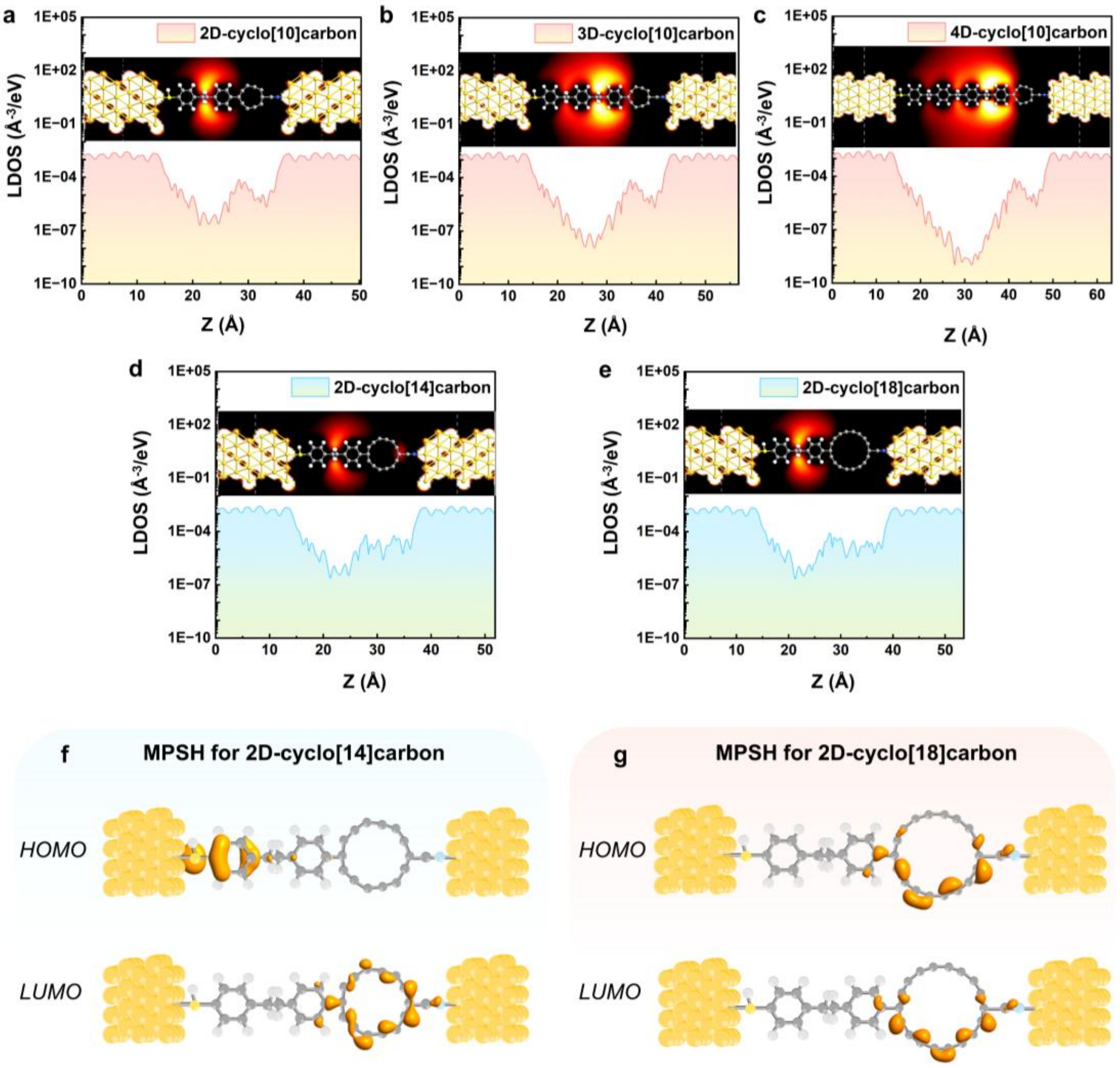

**Figure 5. Electronic states and potential distributions of molecular junctions**. **(a-e)** Local density of states (LDOS) and electrostatic potential differences (EDPs, insets) for 2D-4D cyclo[10]carbon, 2D-cyclo[14]carbon, and 2D-cyclo[18]carbon junctions. **(f, g)** Device MPSH states of the HOMO and LUMO for (f) 2D-cyclo[14]carbon and (g) 2D-cyclo[18]carbon junctions.

**Table 1.** Summary of energy-level alignment and energy gaps for different molecular junctions.

| Molecular junctions | HOMO (eV) | Fermi-level (eV) | LUMO (eV) | Energy Gap (eV) |
|---|---|---|---|---|
| 2D-cyclo[10]carbon | -5.076 | -3.779 | -3.632 | 1.444 |
| 3D-cyclo[10]carbon | -5.073 | -3.779 | -3.631 | 1.442 |
| 4D-cyclo[10]carbon | -5.038 | -3.779 | -3.632 | 1.406 |
| 2D-cyclo[14]carbon | -5.040 | -3.779 | -3.539 | 1.500 |
| 2D-cyclo[18]carbon | -4.668 | -3.779 | -3.593 | 1.075 |

**Table 1** further summarizes the positions of molecular energy levels relative to the electrode Fermi level, along with the effective transport gap after device construction which is obtained from **Fig. S1-2**. The results show that variations in donor length have minimal impact on energy-level alignment and bandgap, indicating that increasing the donor length does not alter the intrinsic electronic structure. In contrast, the cyclo[n]carbon size exerts a pronounced influence on energy-level alignment. Although the LUMO orbital of the 2D-cyclo[10]carbon molecule lies closest to the electrode Fermi level, the 2D-cyclo[18]carbon molecule possesses the smallest bandgap, leading to the highest RR. Nevertheless, it should be emphasized that moderate bandgap narrowing optimizes energy-level alignment and boosts rectification performance, whereas an excessively narrow bandgap increases the DOS near the Fermi level. This enhances off-resonant tunneling under reverse bias and thus reduces the rectification ratio. Furthermore, MPSH analysis further reveals the orbital origin of these trends. While the spatial separation between HOMO and LUMO is preserved for all donor-length variations (**Fig. S6**), increasing cyclo[n]carbon size induces substantial orbital redistribution (**Fig. S7**). In particular, for the 2D-cyclo[18]carbon junction, the HOMO shifts toward the acceptor region, indicating a transition from a conventional D-A transport regime to a (Kornilovitch-Bratkovsky-Williams) KBW-like regime (**Fig. 5**f-g).[42] Despite this evolution, the charge transport remains governed by selective Fermi-level pinning, enabling both high current output and strong rectification without relying on long-range tunneling barriers.

To further assess the generality of the proposed strategy, different electrode materials and acceptor systems are examined (**Fig. S8**-**S10**). When the metallic electrodes such as Ag and Pt are used, the rectification behavior is largely preserved (**Fig. S8**); in contrast, it is noticeably weakened for carbon-based electrodes (**Fig. S9**). Similarly, replacing cyclo[n]carbon with fullerene acceptors does not suppress rectification, but leads to a moderate reduction in rectification performance. This can be attributed to the smaller bandgap of fullerene relative to cyclo[n]carbon (**Fig. S10**). These results highlight that giant rectification is governed by the electrode DOS and the acceptor bandgap. High DOS electrodes stabilize selective Fermi-level pinning, thereby maintaining a robust asymmetric coupling, while wide-bandgap acceptors effectively suppress the reverse-bias charge transport. It is the synergy of these factors that proves essential for achieving high rectification performance.

## Conclusion

This study proposes a design strategy for single-molecule rectifiers based on selective Fermi-level pinning. By synergistically integrating a non-coplanar long-chain donor, a wide-bandgap acceptor, and asymmetric anchoring groups, pronounced spatial separation of FMOs and their selective coupling to electrodes are achieved.

Consequently, LUMO-dominated resonant tunneling is confined to a single bias polarity, enabling a giant RR of up to $10^3$ in representative cyclo[*n*]carbon-based junctions. Importantly, this rectification behavior remains robust against variations in donor and acceptor structures. The turn-on voltage is primarily governed by the built-in electric field, whereas the $RR_{max}$ is determined jointly by the molecular bandgap and energy-level alignment with the electrodes. Generality analysis further reveals that selective Fermi-level pinning is predominantly controlled by the electrode DOS while the upper limit of rectification performance is constrained by the acceptor bandgap. Overall, this work establishes a general physical principle for selective Fermi-level pinning-driven molecular rectification, providing a unified framework and practical guidelines for the rational design of molecular-scale electronic devices.

**Supporting Information**

Supporting Information is available.

**Acknowledgments**

We would like to acknowledge the support from the National Natural Science Foundation of China (NNSFC) (Grant Nos. 52501308), the Key Research and Development Plan of Shandong Province (Grant Nos. 2025CXGC020107 and 2021SFGC1001), the Key Technology Research and Development Program of Shandong Province (No. 2025CXGX010406) and the Natural Science Foundation of Shandong Province (No. ZR2025QC1106). This work is also supported by the Special Funding in the Project of the Taishan Scholar Construction Engineering and the program of Jinan Science and Technology Bureau (2020GXRC019) as well as new material demonstration platform construction project from Ministry of Industry and Information Technology (2020-370104-34-03-043952-01-11).

**Conflict of Interest**

The authors declare no competing financial interest.

**Data Availability Statement**

The data that support the findings of this study are available from the corresponding author upon reasonable request.